\pdfoutput=1 
\documentclass[letterpaper]{article} 
\usepackage[preprint]{aaai2027}  
\usepackage[hyphens]{url}  
\usepackage{graphicx} 
\usepackage{natbib}  
\usepackage{caption} 
\usepackage{booktabs}
\usepackage{array}
\newcolumntype{Y}[1]{>{\raggedright\arraybackslash}p{#1}}

\let\OldTextUnderscore\textunderscore
\renewcommand{\textunderscore}{\OldTextUnderscore\allowbreak}

\title{NetCanvas: Interactive Visual Working Memory for LLM-Based IP Network Fault Localization}
\author{
    Manjing Cai\equalcontrib,
    Chenwei Wu\equalcontrib,
    Qiheng You,
    Weijian Sun,
    Xin Chen
}
\affiliations{
    Huawei Technologies Co., Ltd\\
    caimanjing1@huawei.com
}

\begin{document}

\maketitle

\begin{abstract}
Cross-device IP network fault localization requires large language model (LLM) agents to maintain topology understanding across multi-device, multi-path scenarios. However, mainstream agents keep command-line interface (CLI) output in textual context, so spatial relations such as device adjacency, path forks, and regional hierarchy must be repeatedly reconstructed from scattered text---a failure mode we call \emph{topology amnesia}. We propose NetCanvas, an interactive visual runtime that maintains Interactive Visual Working Memory as probe-synchronized topology images in the agent's context. Through a Probe $\rightarrow$ Ingest $\rightarrow$ Render $\rightarrow$ Act loop, NetCanvas converts CLI observations into an interactive visual topology that updates synchronously with probing. On the CTBench fault localization subset, the interactive visual chain significantly outperforms text-only and structured-graph baselines with fewer average probing steps, and the advantage persists under an additional multimodal LLM backbone. Ablations further attribute the benefit to two sources: probe-synchronized interactive views (vs.\ one-shot static global snapshots) and domain-aligned layouts (vs.\ generic force-directed layouts).
\end{abstract}

\section{Introduction}

Large language model (LLM) agents are being applied to automated network operations (NetOps) and fault localization. Cross-device troubleshooting requires parsing command-line interface (CLI) outputs and maintaining topology understanding across switches, routers, and firewalls. Existing agent paradigms store command outputs, reasoning traces, and intermediate conclusions in textual context, so device adjacency, path forks, and regional hierarchy appear as scattered textual facts. As probing proceeds across multiple diagnostic hops, the model fails to maintain a coherent spatial map. We call this failure mode \emph{topology amnesia}: the agent drops previously inferred upstream connections while parsing downstream device outputs, and must re-infer topological relations at each step. Organizing CLI outputs into a structured JSON graph still forces the model to recover spatial relations from abstract symbols, at a high token cost. This paper targets precisely this representation gap.

\begin{figure*}[t]
\centering
\includegraphics[width=0.92\textwidth]{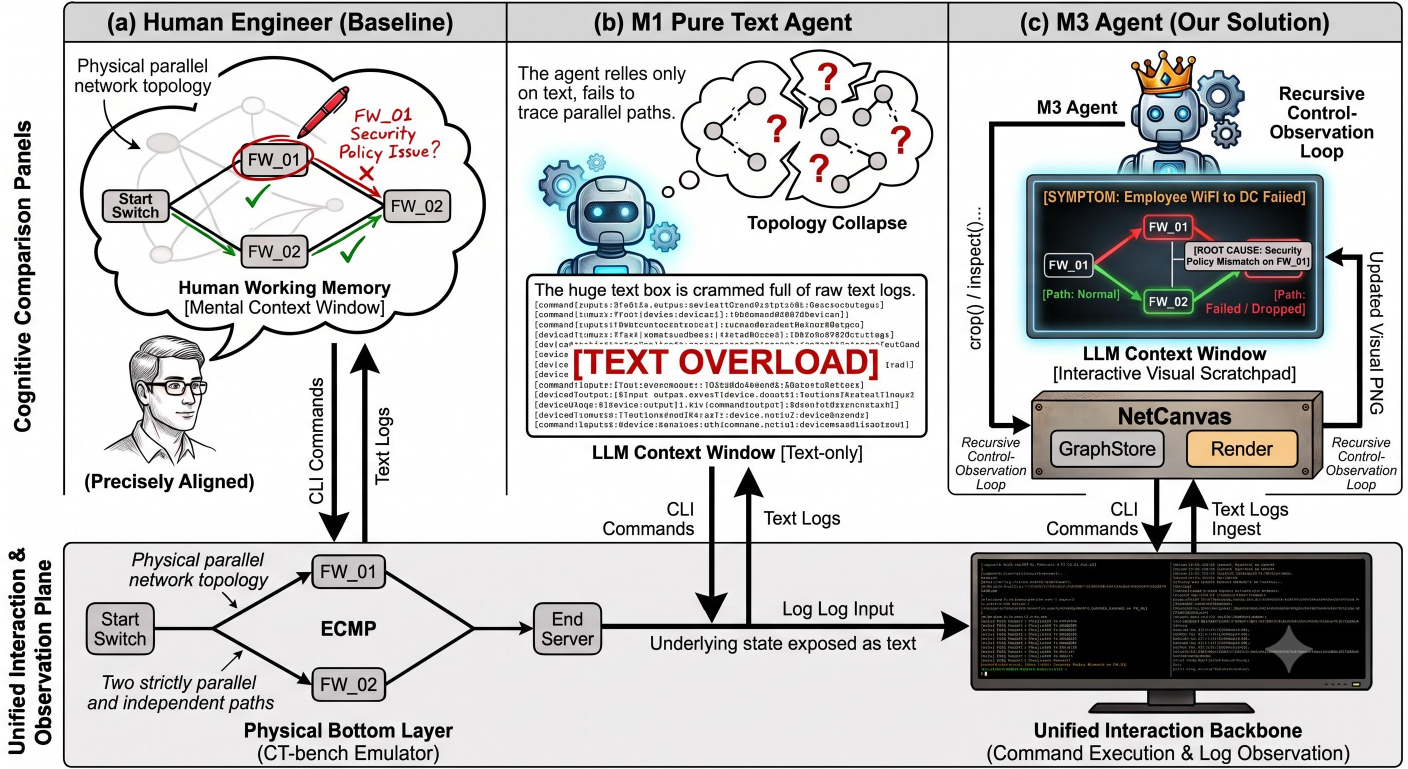}
\caption{Core cognitive comparison. (a)~A human engineer builds a topology diagram in the mind to localize the problem; (b)~a text-only agent (M1) must repeatedly reconstruct device adjacency and path forks from a one-dimensional text log; (c)~an interactive visual agent (M3) continuously preserves spatial structure in context through a topology image synchronized with probing.}
\label{fig:teaser}
\end{figure*}

In real network troubleshooting, engineers commonly rely on topology diagrams for path tracing, local zooming, and device placement confirmation; such diagrams function as external working memory. Motivated by this observation, we propose NetCanvas, an interactive visual runtime that maintains a probe-synchronized topology view $W_t$ in the agent's context as interactive topology images. NetCanvas executes a Probe $\rightarrow$ Ingest $\rightarrow$ Render $\rightarrow$ Act loop that converts CLI observations into an updatable visual representation. Beyond this conversion, the agent navigates $W_t$ on demand---local zoom and crop, path highlighting, and attribute inspection---to obtain local slices of the backend graph state $G_t$. Figure~\ref{fig:teaser} illustrates the comparison: a text-only agent must recover topological relations from a log sequence, whereas an interactive visual agent accesses spatial structure through a topology image synchronized with probing.

We evaluate NetCanvas on the 66-question fault localization subset of CTBench~\cite{ctbench}. The main contributions of this paper are as follows:

\begin{enumerate}
\item \textbf{The NetCanvas framework.} We design an interactive visual runtime for LLM-based IP network fault localization. It organizes the intermediate state in the agent's context as Interactive Visual Working Memory---probeable topology images that replace text-only state tracking---providing a reusable design pattern for NetOps agents.
\item \textbf{Effectiveness of NetCanvas on CTBench.} Without preloading physical topology, the interactive visual chain reaches 54.5\% pass@66, outperforming the text-only baseline (30.3\%) and the structured-graph baseline (37.9\%) with fewer average probing steps; with physical topology preload, it further reaches 63.6\%.
\item \textbf{Design choices that drive the gains.} Controlled ablations show that a probe-synchronized interactive view outperforms a one-shot static global snapshot, and that a domain-aligned layout outperforms a generic force-directed layout. An R-T-H attribution analysis further identifies where visual state tracking improves probing-path coverage and where complex-protocol cases remain bounded by ingest and reasoning bottlenecks (Section~\ref{sec:attrib}).
\end{enumerate}

The paper is organized as follows. Section~\ref{sec:related} reviews related work; Section~\ref{sec:problem} states the problem formulation; Section~\ref{sec:method} presents our method; Section~\ref{sec:exp} reports the experimental setup, evaluation results, and analysis; Section~\ref{sec:conclusion} concludes.

\section{Related Work}
\label{sec:related}

The background of this work is closely related to five lines of research: NetOps agents and benchmarks, tool-augmented agents, agent working memory (text/graph/vision), visual reasoning in multimodal large models (Thinking with Images), and graph layout with domain priors. We review each below.

\textbf{NetOps agents and topology visualization.} Recent benchmarks evaluate various aspects of network automation. NetArena provides a dynamic benchmark generation framework for network automation agents, measuring correctness, safety, and latency on dynamically generated workloads such as routing configuration, data-center planning, and Kubernetes network-policy troubleshooting~\cite{netarena}. Similarly, NetAgentBench offers a state-centric, FSM-formalized dynamic benchmark for multi-turn network configuration, requiring agents to progressively complete routing protocol, connectivity, and policy-related tasks through continuous command interaction~\cite{netagentbench}. In addition, Cornetto benchmarks LLM-driven network configuration repair at scale, requiring models to localize misconfigurations and propose fixes that are verified for functional correctness without introducing new regressions~\cite{cornetto}, while the IETF NetConfBench draft targets intent-driven network configuration, requiring agents to generate and deploy device configurations from user intent and verify them via command matching and test cases~\cite{netconfbench}. These task types are mostly network configuration or network configuration repair, rather than the IP network fault localization this paper focuses on. On the other hand, network fault localization itself relies heavily on a continuously updated topology and path state: early IP topology-discovery research emphasized the importance of automatically constructing network topology for network management~\cite{topodiscovery}, and the OSPF link-state database also reflects the dependence of routing decisions on a global topology view~\cite{rfc2328}. Commercial observability platforms, such as SolarWinds~\cite{solarwinds}, ThousandEyes~\cite{thousandeyes}, and Kentik~\cite{kentik}, also commonly provide topology or path visualization aimed at human engineers. However, when such topology state is no longer presented in a human UI but is instead turned into the agent's working memory, how does its representation form affect troubleshooting effectiveness? Existing research has not sufficiently explored this. To this end, this paper adopts the 66-question fault localization subset of CTBench~\cite{ctbench} as the evaluation basis and, under a unified runtime framework, focuses on the impact of different combinations of ``working memory representation plus interaction protocol'' on diagnostic success rate.

\textbf{Tool-augmented agents.} ReAct~\cite{react} and Toolformer~\cite{toolformer} organize reasoning, acting, and tool use into a general LLM agent paradigm. Our M-series arms are likewise built on an observe--reason--act closed loop and share CLI as the underlying source of network observation; however, the working memory form and tool list exposed to the agent differ across configurations: M1 mainly relies on the textual context of CLI output, M2 reads a structured graph view (JSON subgraphs and graph query results) after ingesting CLI into the graph, and M3 accesses the same backend graph state through visual rendering, path highlighting, and node/edge inspection tools. Accordingly, this paper holds the backbone model, scheduling and scoring rules, and CLI data source fixed, and examines the joint effect of working-memory representation and interaction protocol on diagnostic reasoning.

\textbf{Agent working memory: text, graph, and vision.} MemGPT~\cite{memgpt} introduces virtual context management inspired by OS memory paging to move data between the main context and external storage, but its managed representation remains textual. Recent graph-based memory research~\cite{graphmem} organizes interaction history as external memory: AriGraph~\cite{arigraph} builds and updates a memory graph that integrates semantic and episodic memories during exploration; architectures such as Zep/Graphiti~\cite{zep} persist interaction history into a temporal knowledge graph queryable through node traversal. To reduce token cost over long horizons, OCR-Memory~\cite{ocrmemory} encodes interaction trajectories as images with visual anchors and retrieves verbatim evidence via a locate-and-transcribe mechanism; MemOCR~\cite{memocr} renders structured rich-text memory into layout-aware images with adaptive information density for budget-aware long-horizon reasoning. Unlike the above methods aimed at compressing long-term multi-turn history, this paper focuses on how, within the short-horizon context of a single troubleshooting task, the continuously accumulated topology state should be extracted into an immediate working memory (such as a structured JSON view or an interactive topology image).

\textbf{Visual reasoning and domain interfaces.} In multimodal reasoning, \citet{thinkingimages} survey the Thinking with Images paradigm, in which models use visual marks and tools rather than text-only chains of thought. Building on this line, OpenThinkIMG~\cite{openthinkimg} provides an open framework for tool-augmented LVLM training with visual tool reinforcement learning, and V-Thinker~\cite{vthinker} trains general-purpose agents for interactive, vision-centric thinking via reinforcement learning. However, the above works mainly focus on pixel-level understanding and editing of static raw images. In contrast, images in the network-troubleshooting setting are not raw environmental inputs, but views dynamically rendered by the system based on underlying structured data (GraphStore). The agent does not directly edit pixels; instead, it requests the backend to redraw and highlight through semantic tool instructions such as \texttt{render\_topology(preset="focus")} and \texttt{trace\_route\_path}. This constitutes a domain visual interface over structured state.

\textbf{Graph layout and visual-model reasoning.} Studies such as Graph to Visual and Textual Integration (GITA)~\cite{gita} show that projecting graph structure into visual space---alongside textual graph representations---affects VLM spatial reasoning performance. Building on this line, DynamicGTR~\cite{dynamicgtr} further shows that VLMs exhibit query-dependent preferences over visual and textual graph topology representations, and that dynamically selecting the representation can improve zero-shot graph QA. For network troubleshooting, layout is not only an aesthetic matter; it also encodes engineering semantics such as campus/WAN, core/access, and security domains. Through the comparison between M3 and M3g---under a shared GraphStore and a shared visual interaction protocol, replacing only domain layout coordinates with a generic force-directed layout implemented via NetworkX's \texttt{spring\_layout}~\cite{fruchterman1991,hagberg2008} (Fruchterman--Reingold placement)---this paper tests the independent effect of the domain layout prior on visual reasoning.

\section{Problem Formulation}
\label{sec:problem}

We model cross-device IP network fault localization as a sequential decision-making and information-gathering process based on an implicit graph. This is consistent with the ``observation--model--explanation'' framework of model-based diagnosis (MBD)~\cite{reiter}, but the NetOps setting has stronger spatial structure characteristics: faults are usually not confined to a single device but propagate across multiple devices along links, routing paths, redundant egresses, or security domains. Therefore, troubleshooting is essentially not single-point log analysis but stepwise exploration over an unknown network state. Each CLI probe reveals only local evidence about interfaces, neighbors, routes, policies, or configuration, and the engineer or agent must stitch these scattered pieces into a local topological understanding usable for root-cause localization.

This process can be formalized as the incremental reconstruction of an unknown implicit graph. Let the true and complete underlying network state be an unknown implicit multigraph $G^{*}=(V^{*},E^{*},P^{*})$, where $V^{*}$ denotes device or interface nodes, $E^{*}$ denotes links, adjacency, or forwarding relations, and $P^{*}$ denotes configuration and runtime attributes of nodes or edges. At step $t$ of the diagnosis, the agent generates a probing action $a_t$ based on its current context, i.e., it executes a CLI command on a target device, and the environment returns a textual observation $o_t$. As probing proceeds, the agent gradually reconstructs a growing local topology state $G_t \subseteq G^{*}$ from the observation sequence $o_1, o_2, \ldots, o_t$.

The problem with the current text-only paradigm is that it flattens multidimensional spatial relations into a one-dimensional log sequence. Device adjacency, path forks, redundant links, and cross-region dependencies are originally structural relations on a graph, but in the agent's context they are scattered across multiple rounds of CLI output. As the number of probing hops increases, the model must repeatedly recover the full view of $G_t$ from fragmented text, and is prone to forgetting adjacency relations, confusing paths, repeating probes, or making wrong attributions. As introduced above, this topology amnesia manifests as the repeated loss and reconstruction of the spatial map that diagnosis depends on.

To relieve this cognitive burden, the backend system needs to structure the textual observation $o_t$ and maintain an explicit $G_t$ in a graph database (corresponding to the M2/M3 families in this paper). However, because network topology and runtime attributes grow rapidly with probing, the full $G_t$ or the raw logs are not suitable for direct input into the limited context of an LLM. Therefore, an intermediate cognitive representation is needed in the agent's context, namely a working memory view $W_t$. This is not the backend graph state $G_t$ itself; rather, it is a local state slice projected by the system to the agent according to the current diagnostic goal.

The ultimate task goal is: within a limited number of probing steps $T$, the agent outputs an accurate root cause based on accumulated evidence. With a shared CLI probing data source, this paper's research question can be stated as: given a continuously growing backend $G_t$, how should the working memory view $W_t$ in the agent's context be presented and interacted with to improve the model's understanding of topological spatial relations, and thereby improve fault localization accuracy? This is embodied in two tightly coupled subproblems:

\begin{enumerate}
\item \textbf{Representation form of local topology.} After each probe updates $G_t$, in what form should the extracted subview $W_t$ be injected into the agent's context? Should it remain stacked text (M1), be organized as a structured graph view (JSON subgraphs and graph query results, M2), or be represented as an interactive topology image (M3), to minimize the model's cost of understanding spatial relations in complex topology?
\item \textbf{Interaction protocol for multi-granularity topology.} Diagnosis requires switching across scales such as global overview, local zoom, and path tracing. Beyond changing the return format, what interaction mechanism is needed to let the agent pull slices of $G_t$ on demand, instead of passively receiving a complete graph that overloads the context with information?
\end{enumerate}

\section{Method}
\label{sec:method}

This section details how NetCanvas addresses the two subproblems raised in Section~\ref{sec:problem}. Section~\ref{sec:arch} first outlines the overall data flow and control flow of the NetCanvas visual interaction architecture; Section~\ref{sec:genmech} introduces its core runtime loop, explaining how the working memory view $W_t$ is generated, updated, and presented to the agent; Section~\ref{sec:protocol} defines the visual interaction toolset, explaining how the agent interacts with $W_t$ through a multi-granularity access protocol; and Section~\ref{sec:mapping} explains how the framework maps to the different baseline configurations in the subsequent experiments.

\subsection{NetCanvas Architecture Overview}
\label{sec:arch}

\begin{figure*}[t]
\centering
\includegraphics[width=0.9\textwidth]{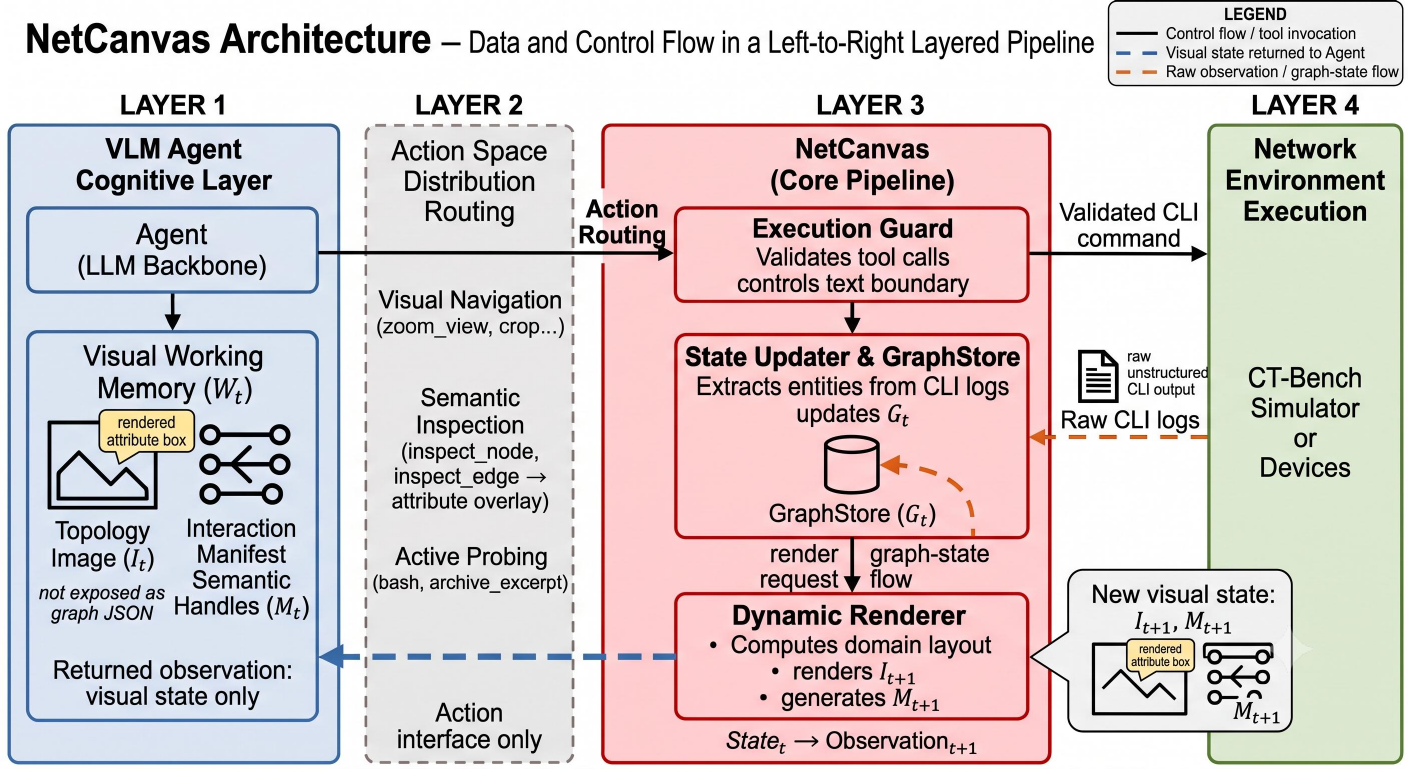}
\caption{NetCanvas Interactive Visual Working Memory architecture. Black solid lines denote control flow, blue dashed lines denote the visual state returned to the agent, and orange dashed lines denote raw observations and backend graph state flow; the agent's context receives only topology images and semantic interaction handles, while the NetCanvas backend handles CLI execution, GraphStore update, and dynamic rendering.}
\label{fig:arch}
\end{figure*}

To enable LLMs to visually access a continuously growing network topology state, we design the NetCanvas runtime. Figure~\ref{fig:arch} presents the control flow and data flow of M3 as a left-to-right layered pipeline. The agent cognitive layer (Layer~1) maintains Interactive Visual Working Memory $W_t$ in context, composed of topology images and semantic interaction handles (Manifest); the action space layer (Layer~2) routes the agent's requests into visual navigation, semantic inspection, and active probing; the backend core engine (Layer~3) is responsible for tool call validation, graph state update, and dynamic rendering; and the network execution environment (Layer~4) executes CLI commands on real devices or a simulator and returns raw textual observations.

\subsection{Generation of Interactive Visual Working Memory}
\label{sec:genmech}

For subproblem~(1)---how the working memory view $W_t$ in the agent's context should be generated and presented---NetCanvas adopts a core runtime loop of Probe $\rightarrow$ Ingest $\rightarrow$ Render $\rightarrow$ Act. This loop converts the textual observations returned by the network environment into an interactive visual topology view and incrementally updates it as probing proceeds.

\begin{enumerate}
\item \textbf{Probe.} The agent issues an active probing request. After validating the request, the NetCanvas backend executes a CLI command on a network device---for example, reading the IP routing table, interface state, or a configuration fragment---and receives the raw textual log returned by the environment.
\item \textbf{Ingest.} The backend engine uses LLM-based information extraction to convert unstructured CLI output into structured objects such as \texttt{RoutingEntry}, \texttt{Interface}, and \texttt{OSPFNeighbor}, and merges them into GraphStore, thereby incrementally updating $G_t$. The updated $G_t$ then feeds the Render stage, which extracts a target subgraph and generates the next version of Interactive Visual Working Memory.
\item \textbf{Render.} The dynamic renderer uses network engineering priors---for example, engineer-curated domain layout coordinates and device connection relations---to extract a target subgraph from $G_t$ and project it into a topology image. Each rendering also generates a Manifest mapping that records entity identifiers, visual positions, and interactive regions. The agent receives the topology image and its corresponding semantic handles; subsequent interactions are issued through semantic identifiers such as devices, links, or paths, and the backend performs positioning, cropping, and redrawing according to the Manifest.
\item \textbf{Act.} The agent receives the updated topology image and semantic interaction handles $W_{t+1}$, reasons based on the current Interactive Visual Working Memory, and selects the next visual-navigation, semantic-inspection, or active-probing action, until it converges on the fault root cause.
\end{enumerate}

\subsection{Multi-Granularity Visual Interaction Protocol}
\label{sec:protocol}

For subproblem~(2)---how the agent interacts with $W_t$ to improve its understanding of spatial relations---NetCanvas designs a multi-granularity visual interaction toolset. Because the M3 agent does not hold the full $G_t$ in context, it must switch among global overview, local zoom, path tracing, and attribute inspection through controlled tools. Each tool call triggers the backend to locate the target element according to the Manifest, render the specified device, link, path, or attribute box in the corresponding view, and return a new visual state $W_{t+1}$ (a topology image and semantic interaction handles). The concrete protocol is shown in Table~\ref{tab:tools}.

\begin{table*}[t]
\centering
\small
\setlength{\tabcolsep}{4pt}
\begin{tabular}{@{}Y{1.85cm}Y{3.35cm}Y{2.65cm}Y{3.25cm}Y{4.15cm}@{}}
\toprule
\textbf{Tool capability} & \textbf{Tool (example call)} & \textbf{Input} & \textbf{Return} & \textbf{Diagnostic role} \\
\midrule
Probe, ingest, and immediate render & \texttt{flush\_and\_render(device, command, question, hops)} & Device, CLI command, question ID, local radius & Planar topology image(s) after ingesting the CLI result, \texttt{manifest\_path}, and a semantic interaction prompt & Ingests CLI into $G_t$ and projects new evidence into Interactive Visual Working Memory \\
\addlinespace
Switch view scope and plane & \texttt{render\_topology(preset, plane, focus, hops)} & \texttt{preset} (overview/focus/path), \texttt{plane}, \texttt{focus/src/dst}, \texttt{hops} & Topology image and semantic interaction handles re-rendered from the accumulated GraphStore & Switches global, local, and path views; toggles physical, logical, and security planes without re-running CLI \\
\addlinespace
Local crop and node quick lookup & \texttt{crop\_node(node, plane, question)} & Node ID, view plane, question ID & 3-hop local topology image, Manifest, and a node attribute box in the footer & Shortcut for inspecting a node neighborhood and its attributes \\
\addlinespace
L3 path tracing & \texttt{trace\_route\_path(src, dst\_ip, dst\_device, question)} & Source device, destination IP/device, question ID & Path description and highlighted topology image from route/next-hop evidence; \texttt{complete:true/false} & Traces the L3 forwarding path, verifying whether the route-hop chain is closed \\
\addlinespace
L2 / physical path enumeration & \texttt{trace\_l2\_paths(src, dst, question)} & Source/destination entity, max paths/hops & Path description and physical-plane topology image from port/LLDP/peer link evidence & Checks port-level or L2 connectivity, complementing L3 trace \\
\addlinespace
Visual object attribute inspection & \texttt{inspect\_visual\_node/edge(manifest, query)} & Latest Manifest and node/edge query string & Matched object highlighted, with an attribute box on the image & Links image objects to underlying structured evidence as visual annotation \\
\bottomrule
\end{tabular}
\caption{Multi-granularity visual interaction tool protocol of NetCanvas (M3). The generic layout mode M3g uses the same interaction logic but replaces domain layout coordinates with a NetworkX spring layout (Fruchterman--Reingold force-directed placement; comparison with M3 in Section~\ref{sec:results}).}
\label{tab:tools}
\end{table*}

\subsection{Mapping the Framework to Experimental Baselines}
\label{sec:mapping}

The architecture above provides a basis for uniformly evaluating different working memory forms. In the controlled experiments in Section~\ref{sec:exp}, the different baseline configurations can be viewed as ablations or substitutions of this framework at different stages:

\begin{itemize}
\item \textbf{Text-only chain (M1).} It skips the Ingest and Render stages, sending the raw textual logs returned by Probe directly to Act as text-only working memory.
\item \textbf{Structured-graph chain (M2).} It retains the Ingest stage to construct $G_t$, but replaces Render with a structured graph view (e.g., JSON subgraphs or graph query results), so as to evaluate the effect of explicit symbolic structure on diagnostic reasoning.
\item \textbf{Interactive visual chain (M3).} It fully executes the Probe $\rightarrow$ Ingest $\rightarrow$ Render $\rightarrow$ Act loop and provides the agent with Interactive Visual Working Memory through the visual interaction toolset.
\end{itemize}

By sharing the underlying CLI Probe and sharing the Ingest pipeline across M2/M3/M3g, the subsequent experiments isolate and quantify the independent contribution of each working memory form to diagnostic success rate; M1 serves as a text-only control that does not explicitly maintain $G_t$.

\section{Experiments}
\label{sec:exp}

\subsection{Experimental Setup}
\label{sec:setup}

\subsubsection{Evaluation dataset.}
The evaluation benchmark is CTBench~\cite{ctbench}, a high-fidelity NetO\&M suite for assessing agentic diagnostic reasoning in realistic communication-technology scenarios. Tasks are curated from industrial maintenance cases through an expert-in-the-loop process and reflect cascading cross-layer faults, partial observability from command outputs, and multi-vendor CLI heterogeneity. In the fault-localization stage, agents must infer a minimal set of normalized root-cause triples (\emph{fault node}, \emph{fault object}, \emph{root cause}) from incomplete evidence, covering route, port, security/NAT policy, Overlay/VPN, and high-availability defects. This paper uses the first 66 fault-localization instances (Q1--Q66) from CTBench, hosted on Hugging Face as \texttt{netop/\allowbreak CTBench}. Table~\ref{tab:clusters} shows the distribution of the 10 fault clusters in this subset.

\begin{table}[t]
\centering
\small
\setlength{\tabcolsep}{4pt}
\begin{tabular}{@{}Y{2.5cm}Y{4.2cm}c@{}}
\toprule
\textbf{Cluster} & \textbf{Mechanism description} & \textbf{\#Q} \\
\midrule
\texttt{l3vpn\_cross\_city} & Cross-city L3VPN path/route anomaly & 15 \\
\texttt{fw01\_policy\_egress} & Firewall policy and egress path issue & 11 \\
\texttt{dual\_fw\_ecmp} & Dual FW, HRP, ECMP, parallel path & 10 \\
\texttt{hq\_core\_stp} & HQ core STP / L2 topology issue & 8 \\
\texttt{edge\_egress} & CSR, NAT, or border-egress issue & 7 \\
\texttt{guest\_cross\_park\_sz} & Cross-campus/city guest access & 6 \\
\texttt{ap\_attach} & AP attachment / dual-point shutdown & 3 \\
\texttt{core\_sz\_stp} & Core/SZ STP / L2 topology & 3 \\
\texttt{core\_vrrp\_local} & Core VRRP / local gateway redundancy & 2 \\
\texttt{fw\_guest\_local} & Guest local / intranet subnet fault & 1 \\
\bottomrule
\end{tabular}
\caption{Fault-cluster distribution (cluster-level pass rates in Table~\ref{tab:cluster}; three-layer attribution of M3 failures in Section~\ref{sec:attrib}).}
\label{tab:clusters}
\end{table}

\subsubsection{Controlled experimental configurations (nine-arm protocol).}
To quantitatively evaluate the impact of different working memory representations, we design nine controlled configurations under the same CTBench CLI data source, unified scheduling and scoring rules, and a unified backbone model (MiniMax-M3, hereafter ``MiniMax''; M1/M2/M3 always refer to working memory configurations). The experiment matrix consists of three working memory representation families, a physical topology preload axis (\texttt{*-0} / \texttt{*-0-0}), and one layout ablation:

\begin{itemize}
\item \textbf{Text-only family (M1 / M1-0 / M1-0-0).} Working memory is the textual context of CLI output. M1 only accumulates probing text; M1-0 additionally preloads a text description of the global physical links; M1-0-0 additionally provides a one-shot global static physical topology image.
\item \textbf{Structured-graph family (M2 / M2-0 / M2-0-0).} Working memory is the structured graph view of the backend GraphStore described in Section~\ref{sec:method} (JSON subgraphs and graph query results). M2-0 preloads a global physical link base graph into GraphStore at the start; M2-0-0 additionally provides a one-shot global static physical topology image.
\item \textbf{Interactive visual family (M3 / M3-0).} Working memory is the interactive topology images and Manifest mapping (generation in Sections~\ref{sec:genmech}--\ref{sec:protocol}). M3-0 preloads a global physical link base graph into GraphStore at the start for subsequent rendering. The visual family has no \texttt{M3-0-0} variant, since the control role of a one-shot static global topology image is already covered by M1-0-0 and M2-0-0.
\item \textbf{Layout ablation baseline (M3g\textsuperscript{\textdagger}).} It shares the same visual interaction protocol as M3, but replaces the domain layout coordinates in the Render stage with a NetworkX spring layout (Fruchterman--Reingold force-directed placement)~\cite{fruchterman1991,hagberg2008}, to isolate the independent gain of the domain layout prior (Section~\ref{sec:results}, RQ3).
\end{itemize}

\subsubsection{Unified protocol and variable control.}
The experiment uses strict unified scheduling and scoring: each question has a maximum of 1~hour; a timeout or failure to output a valid set of root causes is counted as failure. The agent may reflect and retry on tool errors within the time limit, with no additional human intervention. Each (question, arm) combination is independently run up to 3 times. On the information boundary, all \texttt{*-0} and \texttt{*-0-0} variants preload only devices and physical links, without routing tables, ACLs, interface states, or fault labels; logical state must still be obtained through CLI probing. On backend consistency, M2, M3, and M3g share the Ingest pipeline; M3g differs from M3 only by replacing the layout engine; and the text-only baseline M1 does not maintain an explicit graph state $G_t$. In addition, the Manifest is only backend index metadata for rendering positioning, not an independent experimental variable. Therefore, the core comparison is the joint effect of ``working memory representation form plus interaction protocol.''

\begin{figure*}[t]
\centering
\includegraphics[width=0.95\textwidth]{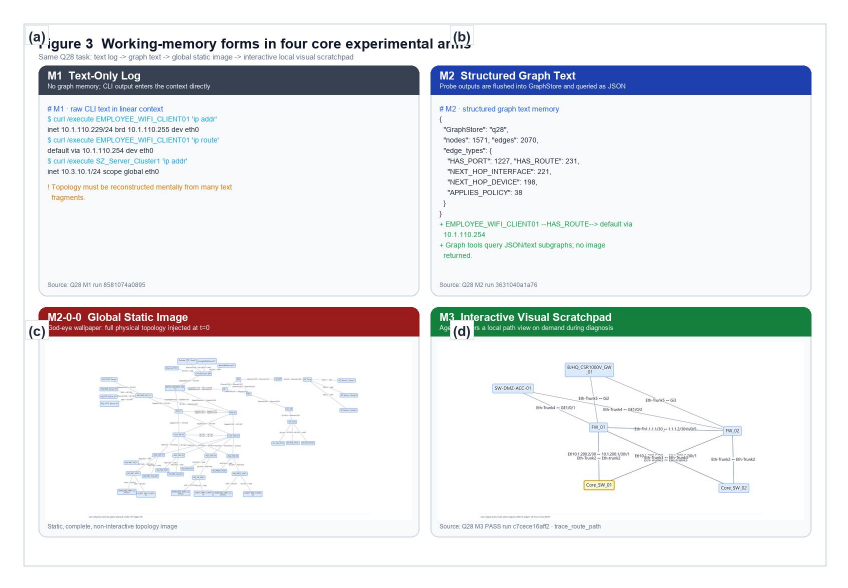}
\caption{Four core working memory forms on Q28 (M1 text / M2 JSON / M2-0-0 static physical snapshot / M3 interactive local rendering).}
\label{fig:memforms}
\end{figure*}

\subsubsection{Evaluation metrics and research questions.}
On the basis of the above configurations, this paper sets three main research questions:

\begin{itemize}
\item \textbf{RQ1 (overall advantage: Interactive Visual Working Memory vs.\ text/structured graph).} Without preloading global topology, can the interactive visual chain (M3) outperform text-only M1 and structured-graph view M2?
\item \textbf{RQ2 (interactivity: dynamic interaction vs.\ static snapshot).} Is the dynamic interactive visual M3 better than configurations that only provide a global static physical topology image in the initial context (M1-0-0 / M2-0-0)? This evaluates the benefit of dynamic on-demand rendering over a one-shot static snapshot.
\item \textbf{RQ3 (layout prior: domain layout vs.\ generic algorithm).} Under the same visual protocol, does the domain layout aligned with network engineering regional semantics (M3) outperform a generic graph layout (M3g\textsuperscript{\textdagger})? This is the closest to a single-variable visual ablation in this paper.
\end{itemize}

Beyond the three main RQs, Section~\ref{sec:aux} separately reports the marginal effect of physical topology preload (\texttt{*-0}), quantifying the gain of the prior knowledge that engineers usually already have about the physical topology before troubleshooting.

\subsubsection{Metrics and scoring mechanism.}
The primary metric is \textbf{pass@66} (the diagnostic pass rate over the full 66-question set). For each (question, arm) combination, hitting the correct root cause on any run counts as passing that question. Overall pass rates and cost trends are in Table~\ref{tab:overall}; per-question paired conclusions for the core RQs are in Section~\ref{sec:results}; and the fault-cluster breakdown is in Table~\ref{tab:cluster}. Resource cost statistics include the average number of tool steps (Avg Steps) and the average token consumption (Avg Tokens), computed as the arithmetic mean over up to 3 completed runs per question. Token refers to \texttt{total\_tokens} (including input, output, and cache-read input tokens), counted regardless of whether a run succeeds, and serves only as a cost-trend reference. Scoring uses strict set matching (Strict Set Match): the root-cause set is extracted from the agent's final output, normalized, and compared strictly with the expert ground truth, with no partial credit or semantic leniency. Given the moderate size of the 66-question set, we report 95\% Wilson score confidence intervals (CIs) for per-arm pass rates. For paired arm comparisons, we use McNemar's exact test on per-question discordant pairs, which matches the shared-question evaluation design.

\subsection{Overall Results and RQ Analysis}
\label{sec:results}

\subsubsection{Overall performance and paired comparison.}
\begin{figure}[t]
\centering
\includegraphics[width=0.98\columnwidth]{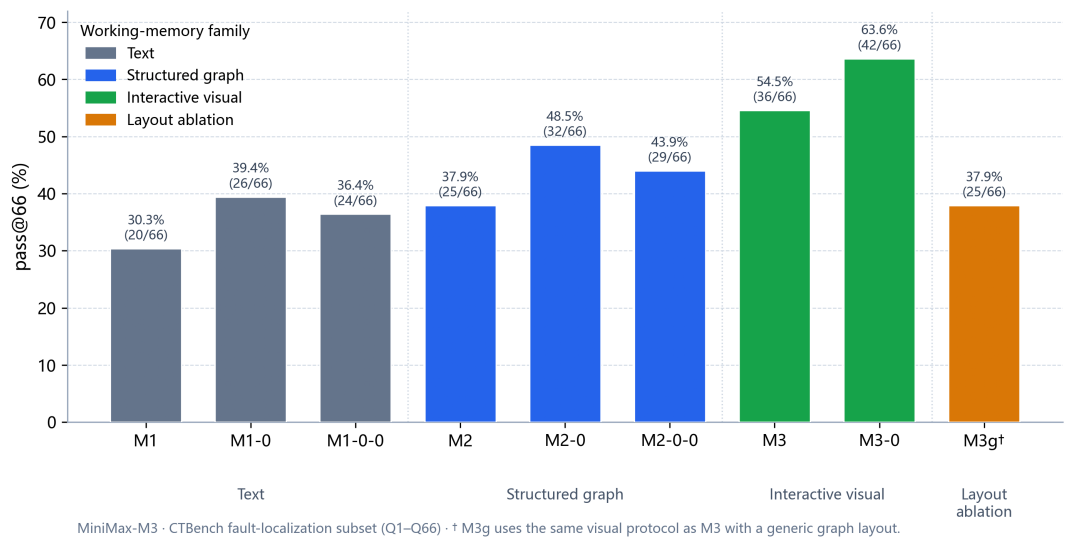}
\caption{pass@66 comparison of the nine working memory configurations on the 66 questions of CTBench. The horizontal axis lists the experimental arms; color distinguishes the working memory families. M3g\textsuperscript{\textdagger} is the layout ablation configuration that shares the visual interaction protocol with M3 but uses a generic graph layout. Full values and resource costs are in Table~\ref{tab:overall}.}
\label{fig:overall}
\end{figure}

We first look at the overall results on the full 66-question set, to compare the overall ranking of the working memory forms (Figure~\ref{fig:overall}, Table~\ref{tab:overall}).

\begin{table*}[t]
\centering
\small
\begin{tabular}{@{}lllccc@{}}
\toprule
\textbf{Family} & \textbf{Arm} & \textbf{Working memory configuration} & \textbf{Pass} & \textbf{Avg Steps} & \textbf{Avg Tokens} \\
\midrule
Text & M1 & Text-only & 20/66 (30.3\%) & 159 & 3,961,465 \\
Text & M1-0 & Text + physical preload & 26/66 (39.4\%) & 159 & 3,804,507 \\
Text & M1-0-0 & Text + static physical topology image & 24/66 (36.4\%) & 145 & 3,229,164 \\
Graph & M2 & Structured graph view & 25/66 (37.9\%) & 217 & 11,998,484 \\
Graph & M2-0 & Structured + physical preload & 32/66 (48.5\%) & 183 & 10,081,813 \\
Graph & M2-0-0 & Structured + static physical topology image & 29/66 (43.9\%) & 182 & 9,232,666 \\
Visual & \textbf{M3} & \textbf{Interactive visual (domain layout)} & \textbf{36/66 (54.5\%)} & 113 & 5,277,545 \\
Visual & M3-0 & Interactive + physical preload & 42/66 (63.6\%) & 107 & 4,598,244 \\
Ablation & M3g\textsuperscript{\textdagger} & Interactive visual (generic layout) & 25/66 (37.9\%) & 131 & 5,809,588 \\
\bottomrule
\end{tabular}
\caption{Overall performance and cost of the nine working memory configurations (pass@66). \textsuperscript{\textdagger}M3g shares probing, ingestion, and interaction protocol with M3; only the Render stage replaces domain layout coordinates with a NetworkX spring layout. Neither arm has physical topology preload.}
\label{tab:overall}
\end{table*}

\begin{figure}[t]
\centering
\includegraphics[width=0.98\columnwidth]{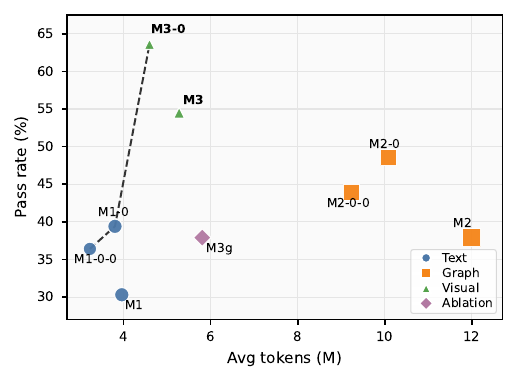}
\caption{Cost--accuracy trade-off of the nine working memory configurations. The horizontal axis reports average token consumption, the vertical axis reports pass rate on CTBench Q1--Q66, and bubble area roughly indicates average tool steps. The dashed line marks the empirical Pareto frontier under lower token cost and higher pass rate.}
\label{fig:costacc}
\end{figure}

Overall, M3 achieves the highest pass rate among no-preload configurations (36/66), and M3-0 with physical topology preload achieves the highest pass rate in the full matrix (42/66). Notably, M3's average number of tool steps is markedly lower than M1 and M2, while its token cost lies between the text-only and structured-graph baselines. Figure~\ref{fig:costacc} further shows that the visual family is not merely trading higher cost for higher accuracy: M3 outperforms all structured-graph variants with substantially fewer tokens, and M3-0 moves the empirical Pareto frontier by improving pass rate while reducing average token consumption relative to M3. In contrast, M3g is dominated by M3, indicating that a generic layout can both reduce diagnostic accuracy and increase exploration cost. Below, we answer the three core research questions in turn, and then break down M3's benefit by fault cluster.

\subsubsection{RQ1: Interactive Visual Working Memory vs.\ text-only and structured-graph baselines.}
Without global topology preload, M3 outperforms both text-only M1 and structured-graph M2 (see Table~\ref{tab:overall}; per-question paired comparisons yield 16 wins / 0 losses against M1 and 14 wins / 3 losses against M2). The 95\% Wilson score CIs are $[20.6\%, 42.2\%]$ for M1, $[27.1\%, 49.9\%]$ for M2, and $[42.6\%, 66.0\%]$ for M3. Although the per-arm CIs of M2 and M3 partially overlap, the paired per-question test controls for question-level difficulty and shows a significant advantage for M3: McNemar's exact test gives $p<0.001$ against M1 (16/0 discordant pairs) and $p\approx0.013$ against M2 (14/3 discordant pairs). Because M2 and M3 share the backend GraphStore ingest pipeline, this difference cannot be explained solely by backend graph construction; it is mainly attributable to the working memory form accessible to the agent and to the visual interaction protocol. On the other hand, the structured-graph view is not without merit: M2-0 with a precise base graph still reaches 32/66 and remains competitive on some questions. Overall, M3's advantage is concentrated more in path-sensitive subclasses that require dynamic focusing, rather than covering all fault forms.

\subsubsection{RQ2: necessity of interaction (dynamic interaction vs.\ static snapshot).}
\begin{figure}[t]
\centering
\includegraphics[width=0.98\columnwidth]{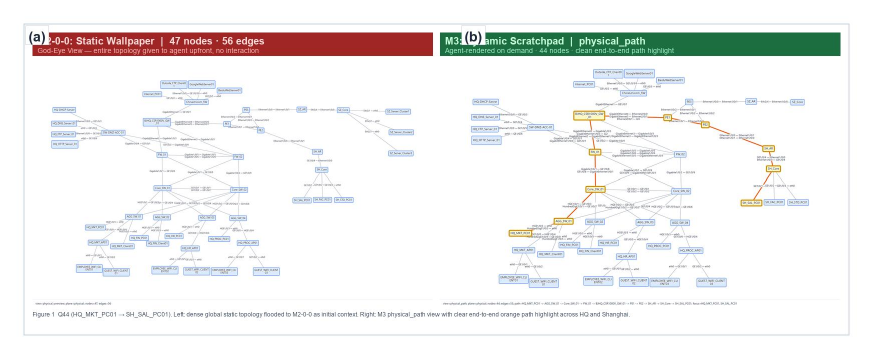}
\caption{A static global diagram provides a one-shot view of physical links, whereas M3 can redraw and highlight the current path on demand during probing. This figure serves as a case illustration of RQ2, complementing the overall cognitive contrast in Figure~\ref{fig:teaser}.}
\label{fig:static}
\end{figure}

Compared with M3's on-demand local rendering, a one-shot global static physical topology image is weaker: M1-0-0 and M2-0-0 reach 24/66 and 29/66 respectively, 12 and 7 questions fewer than M3 (36/66). This confirms that a static global diagram cannot dynamically focus with the probing target and tends to introduce irrelevant visual noise, whereas Interactive Visual Working Memory lets the agent request local views and path highlights on demand, effectively avoiding information overload. As an auxiliary control, M1-0 (26/66) slightly exceeds M1-0-0 (24/66), showing that for the text-only chain a readable structured text prior is usually easier for the model to use than a single static image. This does not mean text-only is the better solution, however: M1-0 still performs markedly below M3 (36/66), which provides a dynamic interactive view (Figure~\ref{fig:static}).

\subsubsection{RQ3: spatial layout constraints (domain prior vs.\ generic algorithm).}
Under a fully shared interactive visual protocol, replacing only the Render coordinates from the domain layout with a NetworkX spring layout (M3g) reduces the diagnostic pass rate from 54.5\% to 37.9\% (M3g underperforms M3 by 11 questions; paired 11 wins / 0 losses, see Table~\ref{tab:overall}). This indicates that, in the NetOps setting of this paper, a layout prior aligned with engineering regional semantics (domain layout prior) is crucial for visual reasoning; an image lacking a clear region hierarchy and device-membership relations instead introduces additional visual distraction.

\subsubsection{Benefit distribution: fault-cluster breakdown.}
To clarify the concrete scenarios of M3's gain, Table~\ref{tab:cluster} further breaks down the pass rates of the core configurations by fault cluster.

\begin{table}[t]
\centering
\small
\setlength{\tabcolsep}{4pt}
\begin{tabular}{@{}Y{2.6cm}cccccc@{}}
\toprule
\textbf{Fault cluster} & \textbf{\#Q} & \textbf{M1} & \textbf{M2} & \textbf{M2-0} & \textbf{M3} & \textbf{M3-0} \\
\midrule
\texttt{fw01\_policy\_egress} & 11 & 90.9 & 90.9 & 100 & 100 & 100 \\
\texttt{dual\_fw\_ecmp} & 10 & 10 & 20 & 90 & 90 & 90 \\
\texttt{hq\_core\_stp} & 8 & 50 & 75 & 87.5 & 87.5 & 100 \\
\texttt{guest\_cross\_park\_sz} & 6 & 0 & 16.7 & 0 & 16.7 & 83.3 \\
\texttt{l3vpn\_cross\_city} & 15 & 0 & 0 & 0 & 6.7 & 13.3 \\
\texttt{ap\_attach}\,+\,\texttt{core\_sz\_stp} & 6 & 0 & 0 & 0 & 0 & 0 \\
\bottomrule
\end{tabular}
\caption{Pass-rate distribution (\%) of selected core configurations by fault cluster.}
\label{tab:cluster}
\end{table}

The data shows that short-path policy tasks (\texttt{fw01\_policy\_egress}) are relatively easy for all baselines, whereas the parallel-path-heavy \texttt{dual\_fw\_ecmp} clearly benefits from physical topology preload or a visual form (the unpreloaded structured-graph M2 gains little here). Cross-city scenarios (\texttt{l3vpn\_cross\_city}) remain difficult across all configurations, meaning that the core bottleneck of this task type goes beyond the working memory form itself; the specific limiting factors are analyzed in depth via the R-T-H framework in Section~\ref{sec:attrib}.

\subsection{Auxiliary Controls and Cross-Model Generalization}
\label{sec:aux}

\subsubsection{Auxiliary control: marginal effect of physical topology preload.}
The \texttt{*-0} variants preload the global physical topology at the start. This section quantifies the marginal gain of this preload across representation families, to judge whether M3's advantage mainly comes from preloading rather than from the working memory form itself (Table~\ref{tab:preload}).

\begin{table}[t]
\centering
\small
\setlength{\tabcolsep}{4pt}
\begin{tabular}{@{}lcccc@{}}
\toprule
\textbf{Family} & \textbf{No prel.} & \textbf{Phys. prel.} & \textbf{Gain} & \textbf{W/T/L} \\
\midrule
Text (M1$\to$M1-0) & 20/66 & 26/66 & $+6$ & 8 / 56 / 2 \\
Graph (M2$\to$M2-0) & 25/66 & 32/66 & $+7$ & 10 / 53 / 3 \\
Visual (M3$\to$M3-0) & 36/66 & 42/66 & $+6$ & 8 / 56 / 2 \\
\bottomrule
\end{tabular}
\caption{Marginal-gain analysis of physical topology preload.}
\label{tab:preload}
\end{table}

The three families gain similarly from preload (6--7 additional correct answers per family; Table~\ref{tab:preload}), indicating that prior knowledge of physical links helps all representation forms uniformly, consistent with the fact that engineers usually have a physical topology prior in real scenarios. With preload added, all families benefit consistently, and M3-0 (42/66) achieves the highest pass rate in the full matrix. On the other hand, preload is not monotonically beneficial: in Q14, M3 passes while M3-0 fails (Appendix~A), suggesting that a global base graph can occasionally induce premature convergence of exploration when the model's reasoning capability is insufficient.

\subsubsection{Cross-model generalization check.}
To test whether the advantage of M3 over M1 depends on a specific backbone, we rerun the M1 and M3 arms with an additional multimodal LLM, keeping the NetCanvas backend, scoring, and CTBench Q1--Q66 unchanged (Table~\ref{tab:xmodel}).

\begin{table}[t]
\centering
\small
\setlength{\tabcolsep}{5pt}
\begin{tabular}{@{}lccc@{}}
\toprule
\textbf{Backbone} & \textbf{M1 (text)} & \textbf{M3 (visual)} & \textbf{$\Delta$ Pass} \\
\midrule
MiniMax\textsuperscript{\textdagger} & 20/66 (30.3\%) & 36/66 (54.5\%) & $+16$ \\
Second backbone & 25/66 (37.9\%) & 39/66 (59.1\%) & $+14$ \\
\bottomrule
\end{tabular}
\caption{Cross-model generalization test on an additional multimodal LLM.}
\label{tab:xmodel}
\end{table}

On this second backbone, M3 still outperforms M1 ($+14$). Absolute pass rates are higher than MiniMax in both arms (M1: 20/66 to 25/66; M3: 36/66 to 39/66), and the paired trend remains significant for both backbones: MiniMax achieves 16 wins / 0 losses (McNemar's exact test $p<0.001$), while the second backbone achieves 17 wins / 3 losses / 46 ties ($p<0.005$). Notably, although its text-only chain is stronger than MiniMax M1 on this task (25/66 vs.\ 20/66), it still does not catch up with MiniMax's interactive visual chain (25/66 $<$ 36/66), further supporting an independent effect of the working memory form on the outcome. The joint effective discriminative subset including the second backbone is in Appendix~C.

\subsection{Failure Attribution Analysis (R-T-H)}
\label{sec:attrib}

\subsubsection{R-T-H three-layer attribution framework.}
Sections~\ref{sec:results} and~\ref{sec:aux} show that M3 leads on most path-sensitive clusters, but it still fails on 30/66 questions. To identify the root causes of these failures, we classify them into three bottleneck types:

\begin{itemize}
\item \textbf{R (Reasoning).} The evidence has appeared, but the model does not converge to the correct mechanism or a minimal set of root causes (divergent output or redundant root causes); the improvement direction mainly relies on model-side post-training (e.g., SFT or reinforcement learning) to strengthen causal inference over the evidence chain, ensuring that it outputs only a minimal necessary set of root causes rather than a divergent enumeration.
\item \textbf{T (Tool-use).} A key probing action did not occur, or the model failed to continue investigating based on the clues returned by the current tools; the improvement direction likewise emphasizes model training and alignment, so that the model internalizes the network troubleshooting probing process and improves its cooperation with the visual interaction tools.
\item \textbf{H (Harness / Ingest).} The key probing action did occur, but the current ingest module insufficiently parses the CLI output, and the evidence is not written into the graph. Here, the Harness refers to NetCanvas's backend probing, ingestion, and rendering pipeline. Because the backend already uses an LLM for information extraction, this bottleneck exposes the ingest-layer extraction capability of the existing ingest LLM when facing complex network protocols. The improvement direction is to perform dedicated fine-tuning of the underlying ingest LLM, strengthening its long-text parsing and structuring ability, thereby greatly improving the structured evidence coverage of complex protocols (such as VRF/RT and prefix-list policy layers).
\end{itemize}

For the concrete attribution, we review the tool trace, GraphStore update records, and final answer of each M3 failed question, and judge in which stage the failure mainly occurs. If a question exposes multiple problems, we record the primary and secondary causes; this analysis is used to explain the failure boundary and guide improvement, not as a new pass-rate metric.

\subsubsection{Failure distribution and core bottlenecks.}
By failure type, M3's 30 failed questions are mainly concentrated in the H+R and T+R combined bottlenecks (Table~\ref{tab:attribcombo}). A combination denotes a superposition of primary and secondary causes; for example, H+R indicates that an insufficient current evidence layer coexists with the model's difficulty in multi-hop reasoning.

\begin{table}[t]
\centering
\small
\begin{tabular}{@{}lccY{3.4cm}@{}}
\toprule
\textbf{Combo} & \textbf{\#Q} & \textbf{Share} & \textbf{Main meaning} \\
\midrule
H+R & 14 & 46.7\% & Insufficient extraction at evidence layer + limited multi-hop reasoning \\
T+R & 9 & 30.0\% & Incomplete probing coverage + divergent output \\
T+H & 5 & 16.7\% & Key device probing interrupted + insufficient structuring \\
R & 2 & 6.7\% & Evidence present, but extra irrelevant root causes \\
\bottomrule
\end{tabular}
\caption{M3 failure attribution: summary by cause combination ($n=30$ failed questions).}
\label{tab:attribcombo}
\end{table}

By fault cluster, these bottlenecks are not uniformly distributed: path-sensitive tasks mostly benefit from M3, while cross-city VPN, AP attachment, and STP-related tasks remain limited mainly by H/R/T combinations (Table~\ref{tab:attribcluster}).

\begin{table*}[t]
\centering
\small
\setlength{\tabcolsep}{5pt}
\begin{tabular}{@{}Y{2.3cm}p{1.9cm}Y{5.6cm}Y{5.4cm}@{}}
\toprule
\textbf{Fault cluster} & \textbf{M3 fails / Bottleneck} & \textbf{Representative phenomenon} & \textbf{Improvement direction} \\
\midrule
\texttt{dual\_fw\_ecmp} & 1/10, R & Most pass via path probing and HRP visualization; Q12 fails to converge & Improve mechanism reasoning and convergence to a minimal root-cause set \\
\texttt{hq\_core\_stp} & 1/8, R & In Q66, STP evidence appears but answers report an extra root cause & Improve answer convergence and pruning of redundant root causes \\
\texttt{edge\_egress} & 3/7, T/R & Egress path, NAT, and border-device probing coverage unstable & Strengthen path probing strategy and tool call order \\
\texttt{guest\_cross\_park\_sz} & 5/6, T+H & Double-fault questions report only FW; remote BJHQ prefix-list not probed/structured & Improve double-fault checklist; extend prefix-list / route-policy ingest \\
\texttt{l3vpn\_cross\_city} & 14/15, H+R & VRF/RT/cross-city routing evidence hard to extract and include & Extend VPN/VRF evidence layer + stronger multi-hop reasoning \\
\texttt{ap\_attach}+\texttt{core\_sz\_stp} & 6/6, T+R & AP attachment and core STP-related probing coverage insufficient & Strengthen access link probing and STP-related reasoning \\
\bottomrule
\end{tabular}
\caption{M3 failure attribution: summary by fault cluster.}
\label{tab:attribcluster}
\end{table*}

The data shows that M3's main performance gain (yielding 16 wins / 0 losses against M1 in per-question paired comparisons) is concentrated in path-sensitive tasks such as \texttt{dual\_fw\_ecmp} and \texttt{hq\_core\_stp}. The main reason is that Interactive Visual Working Memory and its interaction protocol improve probing path coverage (T): with spatialized local views and path highlighting, the agent decides the next probing direction more effectively, turning CLI lookup into spatialized mechanism verification. However, in higher-order failures such as \texttt{l3vpn\_cross\_city}, even with interactive visualization, all configurations still perform poorly (M3 only 6.7\%), mainly attributed to (H)+(R): the current ingest-layer extraction capability does not yet cover the complex VPN routing evidence layer (H), and the model has limitations in multi-hop causal inference (R).

These attribution results define the scope of the current method. M3 mainly improves probing coverage in path-sensitive tasks; but for R-layer problems where the model still fails to converge after evidence has appeared, H-layer problems where evidence from complex protocols is not sufficiently ingested, and probing omissions in complex scenarios, changing the working memory representation and visual interaction protocol alone is not enough. Corresponding improvement directions are discussed in Section~\ref{sec:limit}.

\subsection{Limitations and Future Work}
\label{sec:limit}

\textbf{Topology type and scale generalization.} The 66 fault localization questions come from the same CTBench network environment, sharing the same enterprise campus / cross-city hub-and-spoke topology. Therefore, the current results mainly show that, in this type of multi-site network, NetCanvas can relieve the burden of path tracing and local topology understanding; they cannot be directly extrapolated to other networking forms, such as spine-leaf data-center networks, larger backbone networks, or higher-density multi-tenant cloud networks. Future work needs to reproduce the experiments under different topology scales, hierarchical structures, and device densities, to test the generalizability of NetCanvas and the applicability boundary of the layout prior.

\textbf{Upper bound of logical reasoning.} For failures involving complex protocols such as \texttt{l3vpn\_cross\_city}, a single visual representation form cannot replace a logical reasoning engine. Section~\ref{sec:attrib} attributes their main cause to the H+R superposition (evidence layer + multi-hop inference), which is not unique to M3. Future work will advance along two paths: (1)~the H layer: targeted strengthening of the ingest model's information extraction capability, extending VRF/RT and policy-layer evidence coverage, and introducing a double-fault probing checklist; (2)~the R/T layer: based on the optimized interaction system, conducting domain post-training of the model to internalize network diagnosis capabilities and improve logical convergence and cooperation with the interaction tools.

\textbf{Generalization validation and scoring dependence.} The cross-model comparison (Section~\ref{sec:aux}) shows that the trend of the visual reasoning gain is consistent (second backbone: M3 39/66 vs.\ M1 25/66), but covers only two backbone models and a single network environment; more complete multi-model, multi-network-scale validation is left to future work. In addition, the results depend on skill design, tool protocol, and scoring criteria; strictly counting formatting errors as failures may underestimate the proportion of ``reasoning is correct but expression is non-compliant'' cases. Although validated in the NetOps setting, the ``Probe $\rightarrow$ Ingest $\rightarrow$ Render $\rightarrow$ Act'' closed-loop design of NetCanvas can in principle transfer to other domains with spatial structure (such as circuit design and software architecture), and its cross-domain generalizability remains to be verified. Per-question pass@66 results are in Appendix~A.

\section{Conclusion}
\label{sec:conclusion}

For cross-device IP network fault localization, this paper proposes NetCanvas, an interactive visual runtime that maintains Interactive Visual Working Memory. We formalize the problem as a choice of working memory representation form and build a runtime that extends the agent's working memory from a textual context to a dynamic visual state updated synchronously with probing.

On the CTBench fault localization subset, the interactive visual runtime substantially improves cross-device fault localization: it raises the diagnostic pass rate to 54.5\%, clearly ahead of a text-only baseline (30.3\%), a structured-graph baseline (37.9\%), and even a text-only agent preloaded with a static physical topology (36.4\%), while using fewer probing steps; preloading the physical topology further lifts it to 63.6\%, and the advantage persists under a different backbone model. Controlled ablations attribute these gains to specific design choices rather than visualization alone: probe-synchronized interactive views outperform one-shot static global snapshots, and a domain-aligned layout is critical for multimodal spatial reasoning. A further attribution analysis indicates that the benefit mainly arises from more effective probing-path coverage, while the remaining failures on complex-protocol cases are bounded by the ingest module's evidence extraction and the model's multi-hop reasoning.

\section*{Acknowledgments}
We thank Xingyu Yan, Rong Liu, Changchang Li, Bowen Liu, Mengjie Zhang, Kun Jiang, Tingting Dai, Dingcheng Shan, Sijie Wu, and Yinhang Jiang for their contributions to the construction, curation, and evaluation support of the related public test set, on which the experiments in this paper were conducted.

\bibliography{references}

\begin{thebibliography}{26}
\providecommand{\natexlab}[1]{#1}

\bibitem[{Anokhin et~al.(2025)Anokhin, Semenov, Sorokin, Evseev, Kravchenko,
  Burtsev, and Burnaev}]{arigraph}
Anokhin, P.; Semenov, N.; Sorokin, A.; Evseev, D.; Kravchenko, A.; Burtsev, M.;
  and Burnaev, E. 2025.
\newblock {AriGraph}: Learning Knowledge Graph World Models with Episodic
  Memory for {LLM} Agents.
\newblock In \emph{Proceedings of the International Joint Conference on
  Artificial Intelligence (IJCAI)}.

\bibitem[{Breitbart et~al.(2004)Breitbart, Garofalakis, Martin, Rastogi,
  Seshadri, and Silberschatz}]{topodiscovery}
Breitbart, Y.; Garofalakis, M.; Martin, C.; Rastogi, R.; Seshadri, S.; and
  Silberschatz, A. 2004.
\newblock Topology Discovery in Heterogeneous {IP} Networks.
\newblock In \emph{Proceedings of IEEE INFOCOM}.

\bibitem[{{Cisco Systems}(2025)}]{thousandeyes}
{Cisco Systems}. 2025.
\newblock {ThousandEyes} Platform Overview.
\newblock \url{https://www.thousandeyes.com/}.
\newblock Accessed: 2026-07-03.

\bibitem[{Cui et~al.(2025)Cui, Liu, Xie, and Du}]{netconfbench}
Cui, Y.; Liu, C.; Xie, X.; and Du, C. 2025.
\newblock A Framework to Evaluate {LLM} Agents for Network Configuration.
\newblock IETF Internet-Draft draft-cui-nmrg-llm-benchmark.
\newblock Work in Progress.

\bibitem[{Fruchterman and Reingold(1991)}]{fruchterman1991}
Fruchterman, T. M.~J.; and Reingold, E.~M. 1991.
\newblock Graph Drawing by Force-directed Placement.
\newblock \emph{Software: Practice and Experience}, 21(11): 1129--1164.

\bibitem[{Hagberg, Schult, and Swart(2008)}]{hagberg2008}
Hagberg, A.~A.; Schult, D.~A.; and Swart, P.~J. 2008.
\newblock Exploring Network Structure, Dynamics, and Function using {NetworkX}.
\newblock In \emph{Proceedings of the 7th Python in Science Conference
  (SciPy)}, 11--15.

\bibitem[{{Kentik}(2025)}]{kentik}
{Kentik}. 2025.
\newblock Network Observability Platform.
\newblock \url{https://www.kentik.com/platform/}.
\newblock Accessed: 2026-07-03.

\bibitem[{Li et~al.(2026)Li, Zhang, Yang, Qu, Xu, Yang, Ding, and
  Ngai}]{ocrmemory}
Li, J.; Zhang, Y.; Yang, X.; Qu, J.; Xu, J.; Yang, S.; Ding, J.; and Ngai, E.
  C.-H. 2026.
\newblock {OCR-Memory}: Optical Context Retrieval for Long-Horizon Agent
  Memory.
\newblock arXiv:2604.26622.

\bibitem[{Moy(1998)}]{rfc2328}
Moy, J. 1998.
\newblock {RFC} 2328: {OSPF} Version 2.
\newblock IETF Request for Comments.

\bibitem[{Packer et~al.(2023)Packer, Wooders, Lin, Fang, Patil, Stoica, and
  Gonzalez}]{memgpt}
Packer, C.; Wooders, S.; Lin, K.; Fang, V.; Patil, S.~G.; Stoica, I.; and
  Gonzalez, J.~E. 2023.
\newblock {MemGPT}: Towards {LLMs} as Operating Systems.
\newblock arXiv:2310.08560.

\bibitem[{Protogeros et~al.(2026)Protogeros, Asadli, Hoffman, and
  Vanbever}]{cornetto}
Protogeros, I.; Asadli, R.; Hoffman, B.; and Vanbever, L. 2026.
\newblock Benchmarking {LLM}-Driven Network Configuration Repair.
\newblock arXiv:2604.22513.

\bibitem[{Qiao et~al.(2025)Qiao, Tan, Yang, Dong, Yang, Lang, Wan, Wang, Xu,
  and Yang}]{vthinker}
Qiao, R.; Tan, Q.; Yang, M.; Dong, G.; Yang, P.; Lang, S.; Wan, E.; Wang, X.;
  Xu, Y.; and Yang, L. 2025.
\newblock {V-Thinker}: Interactive Thinking with Images.
\newblock arXiv:2511.04460.

\bibitem[{Rasmussen et~al.(2025)Rasmussen, Paliychuk, Beauvais, Ryan, and
  Chalef}]{zep}
Rasmussen, P.; Paliychuk, P.; Beauvais, T.; Ryan, J.; and Chalef, D. 2025.
\newblock {Zep}: A Temporal Knowledge Graph Architecture for Agent Memory.
\newblock arXiv:2501.13956.

\bibitem[{Reiter(1987)}]{reiter}
Reiter, R. 1987.
\newblock A Theory of Diagnosis from First Principles.
\newblock \emph{Artificial Intelligence}, 32(1): 57--95.

\bibitem[{Schick et~al.(2023)Schick, Dwivedi-Yu, Dess{\`i}, Raileanu, Lomeli,
  Hambro, Zettlemoyer, Cancedda, and Scialom}]{toolformer}
Schick, T.; Dwivedi-Yu, J.; Dess{\`i}, R.; Raileanu, R.; Lomeli, M.; Hambro,
  E.; Zettlemoyer, L.; Cancedda, N.; and Scialom, T. 2023.
\newblock Toolformer: Language Models Can Teach Themselves to Use Tools.
\newblock arXiv:2302.04761.

\bibitem[{Shi et~al.(2026)Shi, Liu, Yang, Mao, Chen, Gu, Su, Cai, Wang, and
  Zhang}]{memocr}
Shi, Y.; Liu, S.; Yang, Y.; Mao, W.; Chen, Y.; Gu, Q.; Su, H.; Cai, X.; Wang,
  X.; and Zhang, A. 2026.
\newblock {MemOCR}: Layout-Aware Visual Memory for Efficient Long-Horizon
  Reasoning.
\newblock arXiv:2601.21468.

\bibitem[{{SolarWinds}(2025)}]{solarwinds}
{SolarWinds}. 2025.
\newblock Network Performance Monitor Product Documentation.
\newblock \url{https://documentation.solarwinds.com/en/success_center/npm/}.
\newblock Accessed: 2026-07-03.

\bibitem[{Su et~al.(2025{\natexlab{a}})Su, Li, Song, Hao, Yang, Zhang, Chen,
  Gu, Li, Qu, and Cheng}]{openthinkimg}
Su, Z.; Li, L.; Song, M.; Hao, Y.; Yang, Z.; Zhang, J.; Chen, G.; Gu, J.; Li,
  J.; Qu, X.; and Cheng, Y. 2025{\natexlab{a}}.
\newblock {OpenThinkIMG}: Learning to Think with Images via Visual Tool
  Reinforcement Learning.
\newblock arXiv:2505.08617.

\bibitem[{Su et~al.(2025{\natexlab{b}})Su, Xia, Ma, Qu, Liu, Li, Zeng, and
  Yang}]{thinkingimages}
Su, Z.; Xia, P.; Ma, Y.; Qu, X.; Liu, J.; Li, Y.; Zeng, K.; and Yang, Z.
  2025{\natexlab{b}}.
\newblock Thinking with Images for Multimodal Reasoning: Foundations, Methods,
  and Future Frontiers.
\newblock arXiv:2506.23918.

\bibitem[{Twabi, Ding, and Kondo(2026)}]{netagentbench}
Twabi, A.; Ding, Y.; and Kondo, T. 2026.
\newblock {NetAgentBench}: A State-Centric Benchmark for Evaluating Agentic
  Network Configuration.
\newblock arXiv:2604.09678.

\bibitem[{Wei et~al.(2024)Wei, Fu, Jiang, Zhang, Zeng, Wu, Kwok, and
  Zhang}]{gita}
Wei, Y.; Fu, S.; Jiang, W.; Zhang, Z.; Zeng, Z.; Wu, Q.; Kwok, J.~T.; and
  Zhang, Y. 2024.
\newblock {GITA}: Graph to Visual and Textual Integration for Vision-Language
  Graph Reasoning.
\newblock In \emph{Advances in Neural Information Processing Systems
  (NeurIPS)}.

\bibitem[{Wei et~al.(2026)Wei, Yan, Kang, Chen, Liu, Kwok, and
  Zhang}]{dynamicgtr}
Wei, Y.; Yan, J.; Kang, C.; Chen, Y.; Liu, H.; Kwok, J.; and Zhang, Y. 2026.
\newblock {DynamicGTR}: Leveraging Graph Topology Representation Preferences to
  Boost {VLM} Capabilities on Graph {QAs}.
\newblock In \emph{Proceedings of the IEEE/CVF Conference on Computer Vision
  and Pattern Recognition (CVPR)}.

\bibitem[{Yan et~al.(2026)}]{ctbench}
Yan, X.; et~al. 2026.
\newblock {CTBench}: Benchmarking Agent Intelligence in Communication
  Technology Scenarios.
\newblock \url{https://huggingface.co/datasets/netop/CTBench}.
\newblock Accessed: 2026-07-06.

\bibitem[{Yang et~al.(2026)Yang, Zhou, Xiao, Dong, Zhuang, Zhang, Wang, Hong,
  Yuan, Xiang, Chen, Zhou, Zhang, Liu, Su, Wang, Chang, and Huang}]{graphmem}
Yang, C.; Zhou, C.; Xiao, Y.; Dong, S.; Zhuang, L.; Zhang, Y.; Wang, Z.; Hong,
  Z.; Yuan, Z.; Xiang, Z.; Chen, S.; Zhou, H.; Zhang, Q.; Liu, N.; Su, J.;
  Wang, X.; Chang, Y.; and Huang, X. 2026.
\newblock Graph-based Agent Memory: Taxonomy, Techniques, and Applications.
\newblock arXiv:2602.05665.

\bibitem[{Yao et~al.(2023)Yao, Zhao, Yu, Du, Shafran, Narasimhan, and
  Cao}]{react}
Yao, S.; Zhao, J.; Yu, D.; Du, N.; Shafran, I.; Narasimhan, K.; and Cao, Y.
  2023.
\newblock {ReAct}: Synergizing Reasoning and Acting in Language Models.
\newblock In \emph{Proceedings of the International Conference on Learning
  Representations (ICLR)}.

\bibitem[{Zhou et~al.(2026)Zhou, Ruan, Wang, Fouladi, Yan, Hsieh, and
  Liu}]{netarena}
Zhou, Y.; Ruan, J.; Wang, E.~S.; Fouladi, S.; Yan, F.~Y.; Hsieh, K.; and Liu,
  Z. 2026.
\newblock {NetArena}: Dynamic Benchmarks for {AI} Agents in Network Automation.
\newblock In \emph{Proceedings of the International Conference on Learning
  Representations (ICLR)}.

\end{thebibliography}

\clearpage
\appendix

\section{Per-Question Pass@66 (Q1--Q66 $\times$ Eight-Arm Main Matrix)}
\label{app:perq}

This table covers the per-question results of the three families and eight variants (M1 \ldots M3-0) in Section~\ref{sec:setup}. The overall performance of the layout ablation M3g\textsuperscript{\textdagger} (25/66) is in Table~\ref{tab:overall}; its per-question matrix is not listed separately due to space. Within the same (question, arm), passing on any of up to three runs is recorded as~P. Eight-arm summary pass@66: M1~20, M1-0~26, M1-0-0~24, M2~25, M2-0~32, M2-0-0~29, M3~36, M3-0~42 (consistent with Table~\ref{tab:overall}).

\begin{table*}[t]
\centering
\small
\setlength{\tabcolsep}{3pt}
\begin{tabular}{@{}c*{8}{c}@{\hspace{1.2em}}c*{8}{c}@{}}
\toprule
\textbf{Q} & M1 & M1-0 & M1-0-0 & M2 & M2-0 & M2-0-0 & M3 & M3-0 &
\textbf{Q} & M1 & M1-0 & M1-0-0 & M2 & M2-0 & M2-0-0 & M3 & M3-0 \\
\midrule
1 & P & P & P & P & P & P & P & P & 34 & P & P & P & P & P & P & P & P \\
2 & F & P & P & P & P & P & P & P & 35 & F & F & F & F & F & F & F & F \\
3 & P & P & P & P & P & P & P & P & 36 & F & F & F & F & F & F & F & F \\
4 & F & F & F & F & F & F & F & P & 37 & F & F & F & F & F & F & F & F \\
5 & F & F & F & F & F & F & F & P & 38 & F & F & F & F & F & F & F & F \\
6 & F & P & F & F & F & F & F & P & 39 & F & F & F & F & F & F & F & F \\
7 & F & P & F & F & F & P & P & P & 40 & F & F & F & F & F & F & F & F \\
8 & F & F & F & P & F & F & F & P & 41 & F & F & F & F & F & F & F & F \\
9 & F & P & P & F & F & P & F & F & 42 & F & F & F & F & F & F & F & F \\
10 & P & F & F & P & P & P & P & P & 43 & F & F & F & F & F & F & F & F \\
11 & F & F & P & F & P & P & P & P & 44 & F & F & F & F & F & F & F & F \\
12 & F & F & F & P & P & F & F & P & 45 & F & F & F & F & F & F & F & F \\
13 & F & F & F & F & P & F & P & P & 46 & F & F & F & F & F & F & F & F \\
14 & F & F & P & F & P & F & P & F & 47 & F & F & F & F & F & F & F & F \\
15 & F & F & F & F & F & P & P & P & 48 & F & F & F & F & F & F & F & P \\
16 & F & F & F & F & P & P & P & P & 49 & F & F & F & F & F & F & F & P \\
17 & F & F & F & F & P & P & P & P & 50 & F & F & F & F & F & F & F & F \\
18 & F & P & P & F & P & P & P & P & 51 & F & F & F & F & F & F & F & F \\
19 & F & F & P & F & P & F & P & P & 52 & F & F & F & F & F & F & P & F \\
20 & P & P & P & P & P & P & P & P & 53 & F & F & F & F & F & F & F & F \\
21 & P & P & P & P & P & P & P & P & 54 & F & F & F & F & F & F & F & F \\
22 & P & P & P & P & P & P & P & P & 55 & F & F & F & F & F & F & F & F \\
23 & P & P & P & P & P & P & P & P & 56 & P & P & F & P & P & P & P & P \\
24 & P & P & P & P & P & P & P & P & 57 & P & P & P & P & P & P & P & P \\
25 & P & P & P & F & P & P & P & P & 58 & P & P & F & P & P & P & P & P \\
26 & P & P & P & P & P & P & P & P & 59 & P & F & F & P & P & P & P & P \\
27 & P & P & P & P & P & F & P & P & 60 & F & P & F & F & P & P & P & P \\
28 & P & P & P & P & P & P & P & P & 61 & F & F & F & F & F & F & F & F \\
29 & F & P & P & P & P & P & P & P & 62 & F & F & F & F & F & F & F & F \\
30 & P & P & P & P & P & P & P & P & 63 & F & F & F & F & F & F & F & F \\
31 & P & P & P & P & P & F & P & P & 64 & F & F & F & P & P & P & P & P \\
32 & F & F & F & F & F & F & P & P & 65 & F & F & P & F & P & F & P & P \\
33 & P & P & P & P & F & F & P & P & 66 & F & P & F & P & F & P & F & P \\
\bottomrule
\end{tabular}
\caption{Per-question pass@66 over the eight-arm main matrix. Left block: Q1--Q33; right block: Q34--Q66. P~=~pass, F~=~fail.}
\label{tab:perq}
\end{table*}

\section{CTBench Task Example (Q28)}
\label{app:q28}

\begin{figure}[t]
\centering
\includegraphics[width=0.98\columnwidth]{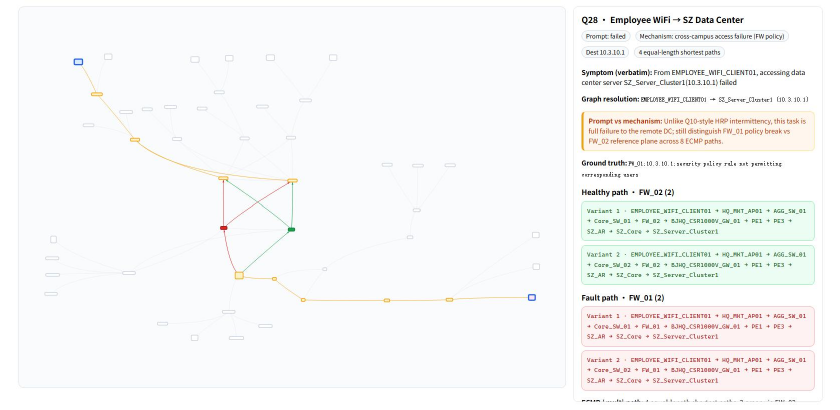}
\caption{Q28 (\texttt{fw01\_policy\_egress} cluster): employee WiFi to SZ data center. The main-text M3 architecture is in Figure~\ref{fig:arch}, and the comparison of working memory forms is in Figure~\ref{fig:memforms}.}
\label{fig:q28}
\end{figure}

\section{Effective Discriminative Subsets (45 / 49 Questions, Pass@66)}
\label{app:subset}

The main tables use the full 66 questions as denominator (Table~\ref{tab:overall}). Some questions still fail under all reported arms, which lowers the pass rate and makes per-question pairing a tie (Win$=$Tie$=$Lose$=$0). This subset is used for the effective discriminative comparison between the second backbone and MiniMax in Section~\ref{sec:aux}. The MiniMax eight-arm effective set (45) requires at least one of the main-matrix eight arms (M1 \ldots M3-0, excluding the M3g ablation) to pass under pass@66; the joint effective set (49) additionally counts questions where 2nd-backbone M1/M3 passes.

\noindent\textbf{All-fail questions (17, outside the joint 49):} Q35--43, Q45--47, Q53--55, Q61, Q63. \textbf{Four questions the 49 set adds over the 45 set:} Q44, Q50, Q51, Q62 (2nd-backbone M3 passes while all MiniMax eight arms fail; M3g is not counted in this definition).

\begin{table}[t]
\centering
\small
\begin{tabular}{@{}lccc@{}}
\toprule
\textbf{Arm} & \textbf{66 questions} & \textbf{49 questions} & \textbf{45 questions} \\
\midrule
M1 & 20/66 & 20/49 (40.8\%) & 20/45 (44.4\%) \\
M2-0 & 32/66 & 32/49 (65.3\%) & 32/45 (71.1\%) \\
M3 & 36/66 & 36/49 (73.5\%) & 36/45 (80.0\%) \\
M3-0 & 42/66 & 42/49 (85.7\%) & 42/45 (93.3\%) \\
2nd-backbone M1 & 25/66 & 25/49 (51.0\%) & 25/45 (55.6\%) \\
2nd-backbone M3 & 39/66 & 39/49 (79.6\%) & 35/45 (77.8\%) \\
\bottomrule
\end{tabular}
\caption{Pass rates over the full set and the two effective discriminative subsets. When discussing only the MiniMax eight arms, the 45-question set is used; when reporting the second backbone at the same time, the 49-question set is used.}
\label{tab:subset}
\end{table}

\end{document}